\documentclass[12pts,fleqn,usenatbib,useAMS]{mnras}

\usepackage{amsmath}
\usepackage{color}
\usepackage{subfigure}
\usepackage{float}

\RequirePackage{fixltx2e}

\def\la{\mathrel{\hbox{\rlap{\hbox{\lower4pt\hbox{$\sim$}}}\hbox{$<$}}}}

\def\ga{\mathrel{\hbox{\rlap{\hbox{\lower4pt\hbox{$\sim$}}}\hbox{$>$}}}}

\usepackage{journal_names}
\usepackage{color}

\usepackage{url}

\newcommand\farcss{\mbox{$.\!\!\!^{\prime\prime}$}}
\def\farcm{\mbox{.\kern -0.5ex\raisebox{.6ex}{\scriptsize$\prime$}}}

\def\farcss{
 \mbox{ 
  \kern  0.13ex. 
   \kern -0.95ex\raisebox{.6ex}{\scriptsize$\prime\prime$}
  \kern -0.1ex
 }
}

\usepackage{graphicx,times}

\usepackage{float}

\title[Pebble isolation mass a proxy for gap opening]{Are the observed gaps in protoplanetary discs caused by growing planets?}

\author[Ndugu et al.]{
N.Ndugu,$^{1,2}$\thanks{E-mail: nndugu@must.ac.ug}
B.Bitsch,$^{2}$
E.Jurua$^{1}$
\\
$^{1}$Department of Physics, Mbarara University of Science and Technology, Mbarara, Uganda\\
$^{2}$Max-Planck-Institut for Astronomy, K{\"o}nigstuhl 17, D-69117 Heidelberg, Germany
}

\pubyear{2019}

\begin{document}
\label{firstpage}
\pagerange{\pageref{firstpage}--\pageref{lastpage}}
\maketitle

\begin{abstract}
{Recent detailed observations of protoplanetary discs revealed a lot of sub-structures which are mostly ring-like. One interpretation is that these rings are caused by growing planets. These potential planets are not yet opening very deep gaps in their discs. These planets instead form small gaps in the discs to generate small pressure bumps exterior to their orbits that stop the inflow of the largest dust particles. In the pebble accretion paradigm, this planetary mass corresponds to the pebble isolation mass, where pebble accretion stops and efficient gas accretion starts. We perform planet population synthesis via pebble and gas accretion including type-I and type-II migration. In the first stage of our simulations, we investigate the conditions necessary for planets to reach the pebble isolation mass and compare their position to the observed gaps. We find that in order to match the gap structures 2000~$\rm M_{\rm E}$ in pebbles is needed, which would be only available for the most metal rich stars. We then follow the evolution of these planets for a few My to compare the resulting population with the observed exoplanet populations. Planet formation in discs with these large amounts of pebbles result in mostly forming gas giants and only very little super-Earths, contradicting observations. This leads to the conclusions that either (i) the observed discs are exceptions, (ii) not all gaps in observed discs are caused by planets or (iii) that we miss some important ingredients in planet formation related to gas accretion and/or planet migration.}
\end{abstract}

\begin{keywords}
accretion, accretion discs, hydrodynamics, protoplanetary discs
\end{keywords}

\section{Introduction}
Recent observations of dust grains in protoplanetary discs have revealed a large level of detailed substructures \citep{ALMAPartnership2015,Andrews2016}. However, up to now it is unclear what mechanism causes the gaps in discs despite of the ever increasing observational evidence for substructures in the discs \citep{Walsh2014,vanderMarel2015,ALMAPartnership2015,Isella2016,Fedele2017,Huang2018,vanderMarel2019}. Proposed mechanisms to explain gaps in protoplanetary discs include dust growth in condensation zones or snow lines \citep{Zhang2015}, zonal flows \citep{Flock2015}, self-induced dust pile-ups \citep{Gonzalez2017}, assemblage sintering \citep{Okuzumi2016}, large scale instabilities \citep{Loren-Aguilar2016}, also notably the  secular gravitational instabilities \citep{Takahashi2016} and, gap opening via clearing by embedded planets \citep{Pinilla2012b,Pinilla2012a,Pinilla2015}. 
% \citep{Lin1979,Kley2012,Baruteau}. 

In this work we will follow the assumption that all gaps in discs are caused by growing planets. Low mass planets only perturb the disc slightly and do not open deep gaps in the disc. However, even low mass planets can perturb the disc large enough to generate a small pressure bump exterior to their orbits. This pressure bump results in reversing the gas-drag such that small dust grains and pebbles are halted in their inward drift \citep{Whipple1972,Paardekooper2006,Pinilla2012a,Morbidelli2012,Lamb2014,Bitsch2018}. This results in an accumulation of particles exterior to the planet which can explain the formation of rings and gaps in the disc observations \citep{Dullemond2018}. Additionally, as the pebbles are trapped exterior to the planet they can not be accreted onto the planet any more. The planet has thus reached pebble isolation mass and can start to accrete gas (e.g. \citealt{Morbidelli2012,Lamb2014,Bitsch2018}). As the pebbles pile up exterior to the planet in rings, a planet that causes such a change in the disc structure must thus be located in a gap of the dust distribution.

Planet population synthesis generates a synthetic planet population by randomizing the initial conditions (e.g. metallicity, planetary embryo starting position, disc lifetime) relevant for planet formation. Previous planet population syntheses considered solid growth via planetesimal accretion \citep{Ida2004,Alibert2009,Mordasini2009}. In these syntheses models, the solid component of the planets is assumed to grow by accreting planetesimals, although numerical simulations show the inefficiency of this process \citep{Tanaka1999,Levison2010}, especially for large planetesimals of several 10km in size. The accretion of small mm-cm sized pebbles, on the other hand, can significantly enhance the growth of planets \citep{Ormel2010,Lambrechts2012,JohansenLacerda2010,Lambrechts2014,Bitsch2015b,Levison2015}. However, most of these simulations have only looked at specific problems like the formation of super-Earths \citep{Lambrechts2019,Izidoro2019,Bitsch2019b} or just gas giants \citep{Levison2015,Chambers2016,Bitsch2019a}.

Recently, pebble-based planet population synthesis approaches appeared in the literature \citep{BitschJohansen2017,Ndugu2018,Chambers2018,Brugger2018,Johansen2019}. These recent models allowed planetary seeds to accrete pebbles until pebble isolation mass, when gas accretion can start. These models also include type-I and type-II migration.

The detailed statistics of observed gaps in discs from the Disk Substructures at High Angular Resolution Project (hereafter, DSHARP) survey \citep{Huang2018} provide us with an excellent statistics of the number and location of rings and gaps in protoplanetary discs. We adopt the pebble-based planet population synthesis approach \citep{BitschJohansen2017,Ndugu2018} to constrain core accretion models to the observed gap statistics \citep{Huang2018}, radial velocity observations of exoplanets \citep{Johnson2010} and to the microlensing observations of exoplanets \citep{Cassan2012}.

This paper is organized as follows: In subsection 2.1 we present the used disc model, explain the planet formation model, and discuss the initial conditions of our simulations in subsection 2.2 and 2.3, respectively. In section 3, we match the synthetic gap opening position data to the observed gap statistics of the DSHARP disc survey. In section 4 we show the match of the synthesized planet population at the end of the disc lifetime with the radial velocity and microlensing data. Section 5 summarizes the main finding in the paper and put them into future investigative perspectives.

\section{Theoretical Models}

\subsection{Disc Model}

Our disc model is derived from the detailed radiative 2D hydrodynamical simulations of \cite{Bitsch2015a} which feature viscous and
stellar heating. The \cite{Bitsch2015a} disc model is based on radially constant accretion rates, where the stellar accretion rate decreases in time for a fixed $\alpha$-viscosity parameter of $0.0054$. Reduction of accretion rates translate into reduced gas surface densities, which in turn reduce viscous heating and thus the disc’s temperature.  The same disc model was used in the pebble-based planet formation simulations \citep{Bitsch2015b,Bitsch2016,Ndugu2018} and in N-body simulations \citep{Izidoro2017,Izidoro2019,Bitsch2019b}. The disc model used in our work has a lifetime which can span all the way up to 10 My as expected by observation \citep{Hartmann,Mamajek2009}.

However, there is now growing evidence that the bulk of discs  are inviscid because the magneto-rotational instability is quenched at all scale-heights in discs \citep{Turner2014}. The accretion is thus thought to be driven by disc winds launched from the surface \citep{Bai2013,Bai2014,Gressel2015,Bai2016,Suzuki2016}. However, there is no clear census yet on how to implement these new features into planet population synthesis models. Recent simulations, for example the planet population synthesis simulations by \citet{Ida2018} and \citet{Johansen2019} adopted similar modifications to their disc models. Following this approach we use $\alpha=0.0054$ to calculate the disc structure, but use a different viscosity parameter to mimic planet migration and the pebble scale height, $\alpha_{\rm mig}$ which can be as low as $0.0001$. Throughout this paper, we note that $\alpha$ is the viscosity that sets the disc's structure (temperature, aspect ratio and gas surface density) and the gas accretion rate onto the planet, while $\alpha_{\rm mig}$ is the bulk viscosity that is reflected in the growth parameters (migration and pebble accretion) of the planetary cores.

\subsection{Planet formation and migration models}

Our planet formation model requires accretion of pebbles onto a core, which eventually triggers runaway gas accretion once the pebble isolation mass is reached and the gaseous envelope contracted. Pebbles with the Stokes number 
\begin{equation}
  \label{eq:Stocknumber}
  \tau_{\rm f} = \frac{{\rho_{\rm {\bullet}}} R}{ { \rho_{\rm g}} H_{\rm g}},
 \end{equation} are accreted onto cores that have reached 
 the pebble transition mass \citep{Lambrechts2012, Lambrechts2014} 
\begin{equation}
\label{eq:transitionmass}
 M_{\rm t} = \sqrt{\frac{1}{3}}\frac{\left({\eta}{\upsilon_{\rm k}}\right)}{G \Omega_{\rm k}}.
\end{equation}
Here, $ \rho_{\rm{\bullet}}$ is the pebble density,  $R$ is the pebble radius, $\rho_{\rm g}$ is the gas density, $ H_{\rm g}$ is the gas scale height, ${\eta}$ is the pressure support, $G$ is the gravitational constant, $ \upsilon_{\rm k}$ is the Keplerian speed and $\Omega_{\rm k}$ the Keplerian frequency.
Initially, the core accretes pebbles in a 3D fashion because the planetary Hill radius is smaller than the pebble scale height \citep{Lambrechts2012,Morbidelli}. As the core agglomerates more pebbles, it becomes more massive and accretes in a 2D manner because $r_{\rm H}>H_{\rm peb}$ \citep{Morbidelli} with a growth rate of 
\begin{equation}
 \label{eq:2Daccretion} 
 \dot{M}_{\rm planet} = 2{\left(\tau_{\rm f}/0.1\right)^{2/3}}{r_{\rm H}}{v_{\rm H}}\Sigma_{\rm peb(r_p)}.
\end{equation} The pebble scale height is given as $H_{\rm peb} = \sqrt{\frac{ \rm \alpha_{mig}}{\tau_{\rm f}}}H_{\rm g}.$                                                                                                                                                                                                                                                         The pebble surface density at the planets position is given by
\begin{equation}
 \Sigma_{\rm peb} = \sqrt{\frac{2\dot{M}_{\rm peb}\Sigma_{\rm g(r_p)}}{\sqrt{3}\pi{\epsilon_{\rm p}}{r_{\rm p}}\upsilon_{\rm k}}},
 \label{eq:pebb}
\end{equation}
where $r_{\rm p}$ denotes the planet's semi-major axis and the pebble flux is given by
\begin{equation}
\label{eq:pebbleflux}
 \dot{M}_{\rm peb} = 2{S_{\rm peb}}{\pi}r_{\rm g}\frac{dr_{\rm g}}{dt}\left(Z_{\rm peb}\Sigma_{\rm g(r_g)}\right).
\end{equation}
Here $Z_{\rm peb}$ is the pebble metallicity, $S_{\rm peb}$ is the pebble flux scaling factor (see below) and $ r_{\rm g}$ the pebble production line given by
\begin{equation}
 r_{\rm g} = \left(\frac{3}{16}\right)^{\frac{1}{3}}\left(G\rm M_{\rm \star}\right)^{\frac{1}{3}}\left({\epsilon_{\rm D}}{Z_{\rm peb}}\right)^{\frac{2}{3}}t^{\frac{2}{3}}
\end{equation}
and 

\begin{equation}
 \frac{dr_{\rm g}}{dt} = \frac{2}{3}\left(\frac{3}{16}\right)^{\frac{1}{3}}\left(G \rm M_{\rm \star}\right)^{\frac{1}{3}}\left({\epsilon_{\rm D}}{Z_{\rm peb}}\right)^{\frac{2}{3}}t^{-\frac{1}{3}},
\end{equation}
where $\rm  M_{\rm \star}$ is the stellar mass, $\epsilon_{\rm p}$ and $ \epsilon_{\rm D}$ are the pebble and dust coagulation efficiency with values 0.5 and 0.05, respectively. This is the standard pebble accretion paradigm \citep{Lambrechts2014} , where $\epsilon_{\rm p}$ and $ \epsilon_{\rm D}$ are values corresponding to detailed coagulation simulations (e.g. \citealt{Birnstiel2011}). Within the framework of \cite{Lambrechts2014}, changes of the accretion rate could only happen due to changes in coagulation efficiency. The here used pebble growth model was described in detail in \citet{Bitsch2015b}. However, \citet{Bitsch2015b} included an unphysical increase in the pebble formation process related to the evolution of the protoplanetary disc structure. This has been discussed in detail in \cite{BitschErratum2018}. We thus here just scale the pebble flux by $S_{\rm peb}$ to allow faster planetary growth, similar to the approach used by \cite{Izidoro2019} and \cite{Bitsch2019a}.

When the planet reaches pebble isolation mass, it accelerates the gas outside its orbit to super-Keplerian velocities, which generate pressure bumps halting pebble accretion, trapping inward moving pebbles and allows the planet to transition into a stage of gas accretion  \citep{Lamb2014}. Through out our simulations, the new pebble isolation mass relation of \cite{Bitsch2018}  given by
\begin{eqnarray}
\label{eq:Iso}
 {M_{\rm iso}}&\approx&25\left(0.34\left(\frac{\log_{10}(0.001)}{\log_{10}(\alpha_{\rm mig})}\right)^4 + 0.66\right)\nonumber \\ &&\left(1 - \left(P_{\rm grad} + 2.5\right)/6\right)\left(\frac{H/r}{0.05}\right){\rm M}_{\rm E}
\end{eqnarray} is used. $P_{\rm grad}$ corresponds to the radial pressure gradient in the unperturbed protoplanetary disc. This clearly shows the importance of the disc structure and viscosities for planetary growth and migration simulations.

At the pebble isolation mass, the gas envelope contracts over a long time while accreting some gas as long as $ \rm M_{\rm env} < \rm M_{\rm core}$ at a rate given by \citep{Pisso,Bitsch2015a}
% \newpage
\begin{eqnarray}
\label{eq:envelope}
 { \dot{M}_{\rm gas}} & = &0.000175f^{-2}\left(\frac{ \kappa_{\rm env}}{1 \rm  cm^{2}/g}\right)^{-1}\left(\frac{ \rho_{\rm C}}{5.5 \rm  g/cm^{3}}\right)^{-\frac{1}{6}} \nonumber \\ && \left(\frac{M_{\rm c}}{\rm M_{\rm E}}\right)^{\frac{11}{3}} \left(\frac{M_{\rm env}}{1 \rm M_{\rm E}}\right)^{-1} \left(\frac{T}{81 \rm K}\right)^{-0.5}\frac{\rm M_{\rm E}}{\rm My} \ ,
\end{eqnarray}
where $f$ is a scaling factor given by 0.2 \citep{Pisso} and $ \kappa_{\rm env} = 0.05 {\rm cm^{2}/g} $ is the planet's envelope opacity very similar to the study by \cite{Movshovitz}. $\rho_{\rm C} $ is the core density and we assumed $\rho_{\rm C}$ = 5.5 $\rm{g/cm^{3}}$.
The planet continues to contract its envelope until $M_{\rm c} = M_{\rm env}$. When $ M_{\rm c} < M_{\rm env}$, rapid gas accretion is triggered and the gas accretion rate is given by the minimum of \citep{Machida}
\begin{equation}
 \dot{M}_{\rm gas,\rm low} = 0.83\Omega_{\rm k}\Sigma_{\rm {g}}H^{2}\left(\frac{r_{\rm H}}{H}\right)^{\frac{9}{2}}
\end{equation}
and 
\begin{equation}
 \dot{M}_{\rm gas,\rm high} = 0.14 \Omega_{\rm k}\Sigma_{\rm g} H^{2}.
\end{equation} 
We note that these gas accretion rates might be too generous and allow the formation of too many gas giants (e.g. \citealt{Brugger2018,Bitsch2019a}), however, until the planets reach pebble isolation mass gas accretion can be neglected. We will investigate different gas accretion rates in an upcoming publication and first tests imply that different gas accretion rates do not influence our results significantly.

Low mass planets radially drift inwards in the type-I fashion where their migration is modeled following the torque formula from \cite{Paardekooper2011}, where the total torque, $ \Gamma_{\rm tot} $ acting on the planet is given by 
\begin{equation}
    \label{eq:tottorque}
   \Gamma_{\rm tot} = \Gamma_{\rm L} + \Gamma_{\rm C}.
\end{equation}
  $ \Gamma_{\rm L}$ and $ \Gamma_{\rm C}$ are the Lindblad and corotation torques, respectively. The Lindblad and corotation torques strongly dependent on the local radial gradients of gas surface density, $ \Sigma_{\rm g} \propto r^{-\lambda} $, temperature $T \propto r^{-\beta} $, and entropy $S \propto r^{-\varepsilon} $, with $ \varepsilon = {\beta} + \left(\gamma - 1.0\right)\lambda $ and $\gamma = 1.4  $ is the adiabatic index. When the planet grows massive enough, it carves gap in the disc \citep{Crida2006} and eventually switches to the slower type - II migration \citep{Baruteau}:
  \begin{equation}
      \tau_{\rm II} = \tau_{\rm \upsilon} \times  max\left(1, \frac{M_{\rm p}}{4\pi \Sigma_{\rm g} r_{\rm p}^2}\right).
      \label{reducedmig}
  \end{equation}
Here, $\tau_{\rm \upsilon}$ is accretion time scale of the protoplanetary disc. Combining the disc model, planet formation and migration models, we synthesize population of planets by sampling the initial conditions. For this, we slove equation 1-13 numerically. This is described in detail in the following subsection.

\subsection{Initial conditions}
Planet populations synthesis requires a randomization of important input parameters of discs and planet formation as these parameters are not a-priori known. These important input conditions are: (i) Disc metallicity and lifetime, which set the disc structure ($\Sigma_{\rm g}$, H/r) and thus influences the growth and migration speeds. The disc structure and its evolution are described in detail in \cite{Bitsch2015a}. The gas surface density follows in the outer parts of the disc a power law with radius of -14/15 and the aspect ratio follows a 2/7 power law in radius. (ii) Implantation time and position which sets the growth and migration trajectories of cores. Our input parameters are sampled in similar manners as in \cite{Ndugu2018} but with the exception of subjecting the disc populations to exponentially decaying lifetimes as predicted by observations (e.g. \citealt{Hartmann,Mamajek2009}). The disc lifetime is exponentially decaying with a half lifetime of 2.5 My, where we fixed the minimal and maximal lifetime of the disc to 0.1 and 10 My, respectively.
Throughout the simulation, our exponentially-decaying surface density prescription is not derived from any underlying physical model but purely empirically derived from observations.
The initial starting locations follow a linear distribution in semi-major axis. We show results for a logarithmic starting configuration in the appendix, but this does not affect our overall conclusions. Our seeds are placed up to 160 au. The starting time of the planetary seeds corresponds to the time when the pebble production line (Eq.6) reaches the starting position plus 100 orbital times, as this corresponds roughly to the timescale of the streaming instability to form planetesimals (e.g.  \citealt{Johansen2015}). We then start our planetary embryos at the pebble transition mass. This approach might not be realisitic, as planetesimals might take a long time to coagulate into planetary embryos at large distances, it thus gives us the maximal possible growth time by pebble accretion for embryos in the outer disc.
% \begin{eqnarray}
% t_{\rm 0} = \frac{r_{0}\times AU}{ \left(\frac{3}{16} \right)^{\frac{1}{3}}\times(GM_{\star})^{1/3}(\epsilon_{D}Z)^{2/3}} + \left(200\pi\sqrt{\frac{(r_{0}\times AU)^{3}}{GM_{\rm \star}}}\right),
% \end{eqnarray}

\section{Match to the DSHARP campaign}

The DSHARP campaign limits the alpha-viscosity parameter within the range of $10^{-6}$ to $10^{-2}$. Our choices of alpha are thus within the limits of the observations. A higher viscosity results in higher pebble isolation masses, resulting in planets reaching the pebble isolation mass only in the inner regions of the disc. Lower viscosities, on the other hand, reduce the pebble isolation mass and allow planets to reach the pebble isolation mass also in the outer parts of the disc.
We generated a population of cores that just reach pebble isolation mass following the initial conditions described in subsection 2.3.  Our approach is the simplistic one-embryo-per-disc. We are aware that single embryo-per disc planet population synthesis models might present an over estimate in the growth efficiency of cores due to the lack of competition for materials and the neglection of eccentricity effects which might affect the accretional efficiency \citep{J2015,LiuOrmel2018,OrmelLiu2018} or the migration rates \citep{BitschKley2010,Cossou2013}. Our result is therefore the upper limit representation of the accretional efficiency of single planets for materials in the discs.

We make the assumption that the gaps in the protoplanetary discs are opened at positions where the planetary embryos have reached pebble isolation mass. To compare our simulations to the observations, we stop planet growth and migration once the planet reaches pebble isolation mass and track at what time the planet reached pebble isolation mass. We follow the evolution of 3000 planetary embryos for each set of simulations, where the positions of the embryos that have reached pebble isolation mass is compared to the position of the gaps seen in observations. The observed gap positions are extracted from Fig. 2 of  \cite{Huang2018} which is normalised and compared to the normalised occurrence of the simulated gap positions as displayed in Fig.1.

\begin{figure}
 \centering
 \includegraphics[width=\columnwidth]{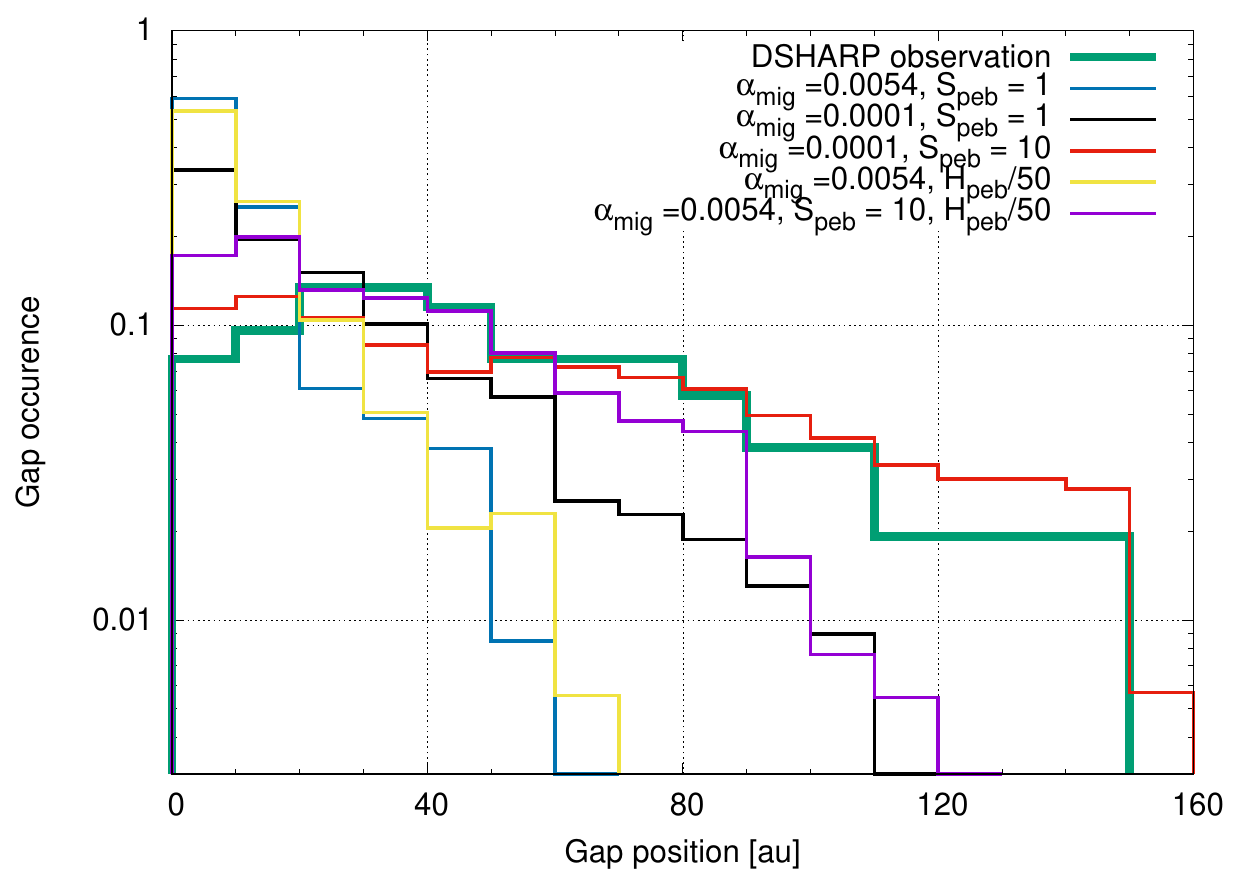}
 \hfill
 \caption{Comparison of the gap occurrence rate found in protoplanetary discs by the DSHARP survey \citep{Huang2018} and positions of planets in our synthesized population that have just reached pebble isolation mass. The planetary growth and migration is stopped when the planets have reached pebble isolation mass, as they start to generate small pressure bumps in the disc that stops the inflow of pebbles and can thus generate rings exterior to the planets position and gaps at the planets position. These simulations clearly show that a very high pebble flux and low pebble scale height (simulations marked with $H_{\rm peb}/50$ feature a large $\alpha_{\rm mig}$ for migration, but a reduction in pebble scale height by a factor of 50 to track the influence of the migration rates on the gap position) is needed to make planets reach the pebble isolation mass at large orbital distances.}
%   \label{fig:1}
\end{figure}

From Fig.1 we see that matching the DSHARP observed gaps requires a very high (and probably unrealistic) pebble flux scaling factor, low alpha for migration and particle scale height. Otherwise the observed gaps at the outskirts of the disc can not be reproduced. For low pebble fluxes and larger alpha parameters, the outer gaps can not be matched, because growth is too slow in the outer disc. Low alpha means slower migration and very thin particle scale height, therefore cores radially migrate slower to the inner discs and transit earlier into the faster 2D pebble accretion fashion and can therefore reach the high pebble isolation mass in the outskirts of the protoplanetary disc earlier. At high alpha specifications, even higher pebble flux scaling factors are required to match the observations of the protoplanetary discs. 

For a planet to reach the pebble isolation mass in the $S_{\rm peb} = 10$ case in the outer disc (red curve in Fig.1) for a disc with [Z/H]=0.5 (see Fig.2), a total of about 2000 Earth masses of pebbles have to pass the planet. This amount of pebbles is about a factor of 10-20 higher than the pebbles trapped in the observed rings from DSHARP survey \citep{Dullemond2018}. In our model, [Z/H]=0.0 corresponds to a dust-to-gas ratio of 0.015, indicating that a disc with [Z/H]=0.5 would feature a dust-to-gas ratio of 0.047. This implies that the gas mass of a disc with [Z/H]=0.5 containing 2000 Earth masses in pebbles corresponds to $\approx$ 0.14 Solar masses, which is on the higher mass ranges of protoplanetary discs. We also note that in order to match the gaps seen by the DSHARP survey our simulations require a pebble production line that extends all the way to  around 300 au. This is clearly beyond the the outer boundaries of the disc according to observations.

Therefore given the limitations of our model, reproducing the observed gaps in the DSHARP survey requires low alpha values, a very extended pebble production line and a pebble scaling factor of around $S_{\rm peb} = 10$ resulting in very large amounts of pebbles needed to form planets at the position of the gaps. 

Setting the pebble scale height, $H_{\rm peb}$ lower by a factor of $50$ but keeping $\alpha_{\rm mig}=0.0054$, we investigate how the match of our simulations to the gaps seen by the DSHARP survey changes. Lowering the pebble scale height helps the growth, but the gaps at the very outskirts of the discs are not matched. This is because although the transition into effective 2D peble accretion is faster, the seeds drift inwards before generating pressure bumps.

\begin{figure}
 \centering
 \includegraphics[width=\columnwidth]{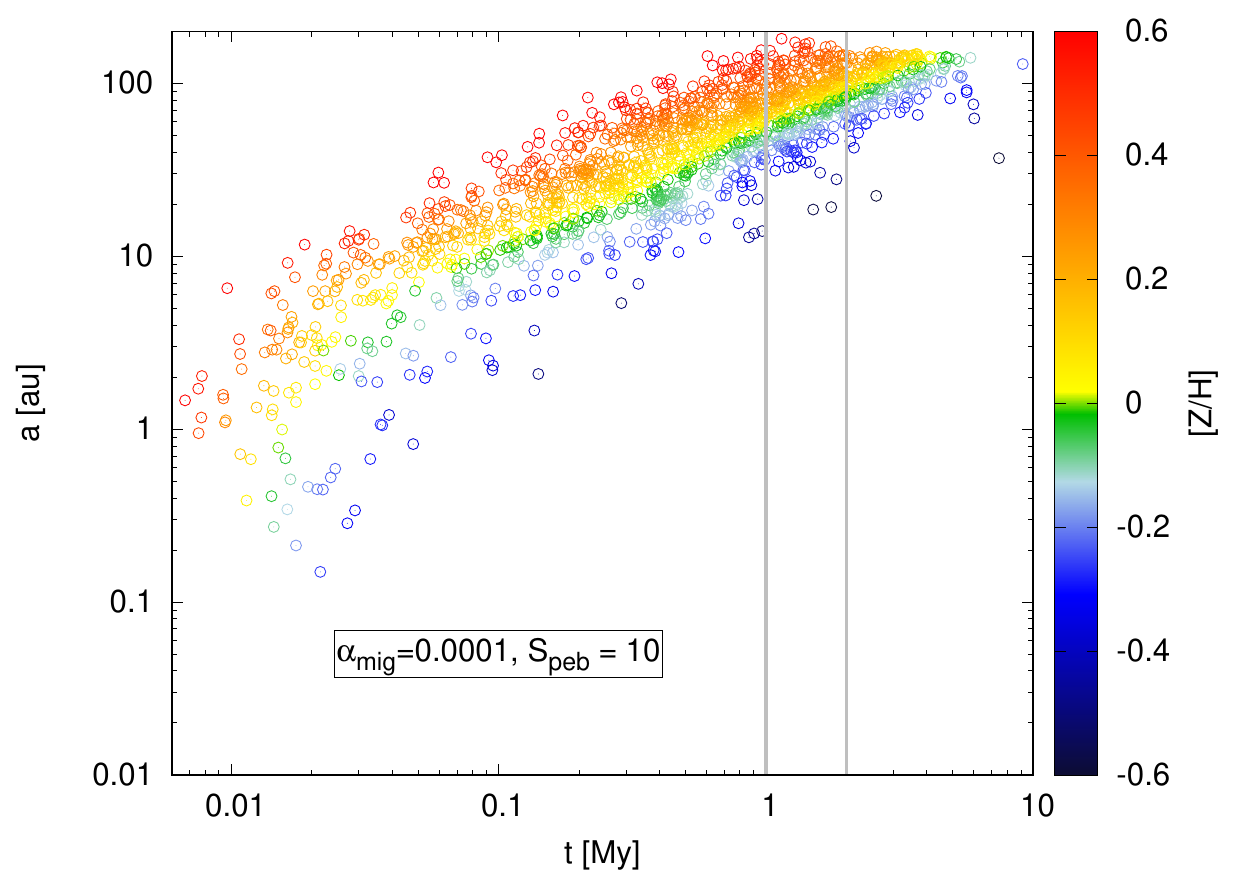}
 \hfill
 \caption{Position and time for planetary embryos to reach pebble isolation mass as a function of metallicity. Different colored dots shows the metallicity of the disc environment from which the core formed. From the plot, it is easily seen that seeds reach pebble isolation with a time spread of 0.01 My to 10 My which is much broader than the disc lifetime specification from DSHARP campaign (1:2 My). Cores that reached pebble isolation mass between the two grey vertical line in the plot fall in the disc lifetime of DSHARP survey, marked by the two vertical grey lines. Cores growing in different disc metallicities reach pebble isolation at different position. For example, to reach pebble isolation mass at the outskirt of the disc at the same time requires higher metallicities than for inner discs.}
%   \label{fig:1}
\end{figure}

We show the time at which planets reach pebble isolation mass as a function of the host star metallicity in Fig.2. This figure shows a much wider spread in time of gap opening (0.01:10 My) than the specifications in the DSHARP campaign (1:2 My).The planets in the inner regions of the disc grow very fast, because there pebbles are produced immediately and the accretion rates are very high, enhancing the growth. Fig.2 tells us that making gaps in the outskirts of the discs require large amount of materials and additionally a large metallicity to achieve efficient planet growth in the outer regions of the protoplanetary discs. Lower metallicity implies that planet formation takes longer than the observed disc ages (1-2 My). This would imply that there is a correlation between host star metallicity and distance of giant planets. The more metal rich a host star is, the farther away giant planets could be located.

Following this time specifications for planetary embryos reaching isolation mass, full simulation runs were executed until the disc reaches its final lifetime of up to 10 My. These simulations were obtained for the model with $\alpha_{\rm mig} = 0.0001$ and $S_{\rm peb} = 1$ and for the model that gives good match to the gap occurrence rates of the DSHARP survey (model with $\alpha_{\rm mig} = 0.0001$ and $S_{\rm peb} = 10$). These models are indicated by the black and the red curve in Fig.1. We show the results of these planet population synthesis models in Fig.3 as function of the disc lifetime and in Fig.4 as a function of host star metallicity. These plots clearly show the importance of disc lifetimes and metallicity in the synthesised planet population simulations. These simulations clearly predict a large population of giant planets that could be seen by direct imaging surveys, for example, the HR8799 system, but rarely by the RV method. The observed gas giants in reality are mainly on eccentric orbits, therefore they most likely underwent gravitational scattering \citep{JuricTremaine2008,Sotiriadis2017}. This means the large population of synthesised giant planets in the direct imaging field of view could be scattered by gravitational interaction which might refill the dearth of the planets in the RV field of view should N-body interaction between the protoplanets been incorporated. However, the planets in our simulation seem to be too massive compared to the observations. The comparison of subsets of the synthesised population of planets to the observational limits of the RV and the microlensing technique is shown in section 4.

\begin{figure}
      % \centering

         \includegraphics[width=\columnwidth, height=1.75in]{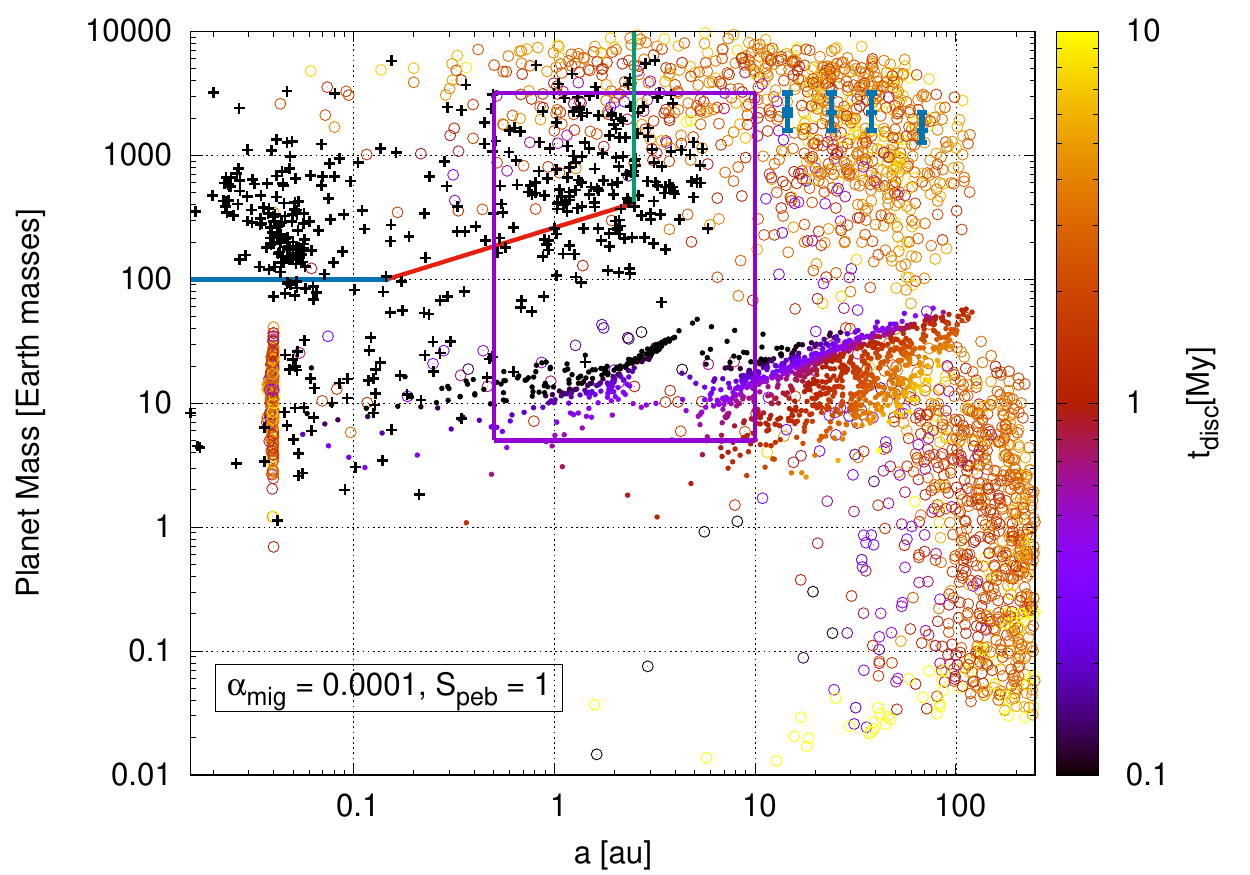}
         \hfill
         \includegraphics[width=\columnwidth, height=1.75in]{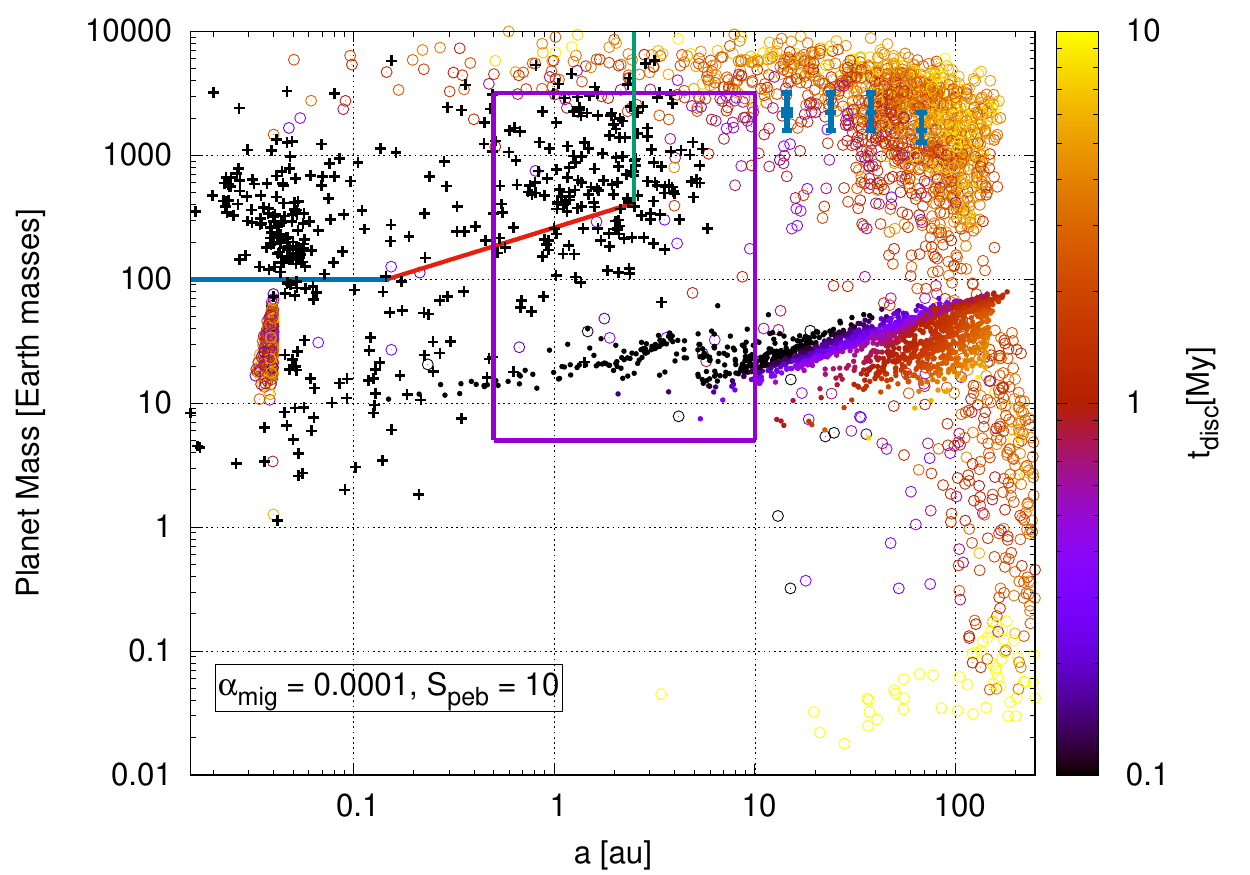}
         \hfill
         \caption{The top plot feature simulations with nominal pebble flux and lower alpha (for migration and pebble scale height) and the bottom plot features results of a simulations with $ S_{\rm peb} = 10 $, which gave the best match to the gap observations (Fig.1). The black crosses are the exoplanet population observed by the RV method taken from the exoplanet.org database.  In all of the plots, the blue crosses are the HR8799 planetary system. The solid dots denote the mass and the position of the planet at pebble isolation mass, while open circles denote the planet population that has been integrated until disc dissipation. The color of each dot shows the time at pebble isolation mass (solid dots) or the disc lifetime (open circles).  The red straight line is the RV detection line for giant exoplanets. The horizontal blue line is the division line for graduating a planet into a gas giant. The green straight line at 2.5 au represents the orbital limit of the RV exoplanet survey of \citet{Johnson2010}, indicating that only gas giants in this upper left quadrant of the plot can be observed by the RV technique. Planets inside the purple rectangle can be observed by the microlensing technique \citep{Cassan2012}.}
%          \label{fig:3}
 \end{figure}

\begin{figure}
        \centering

         \includegraphics[width=\columnwidth, height=1.75in]{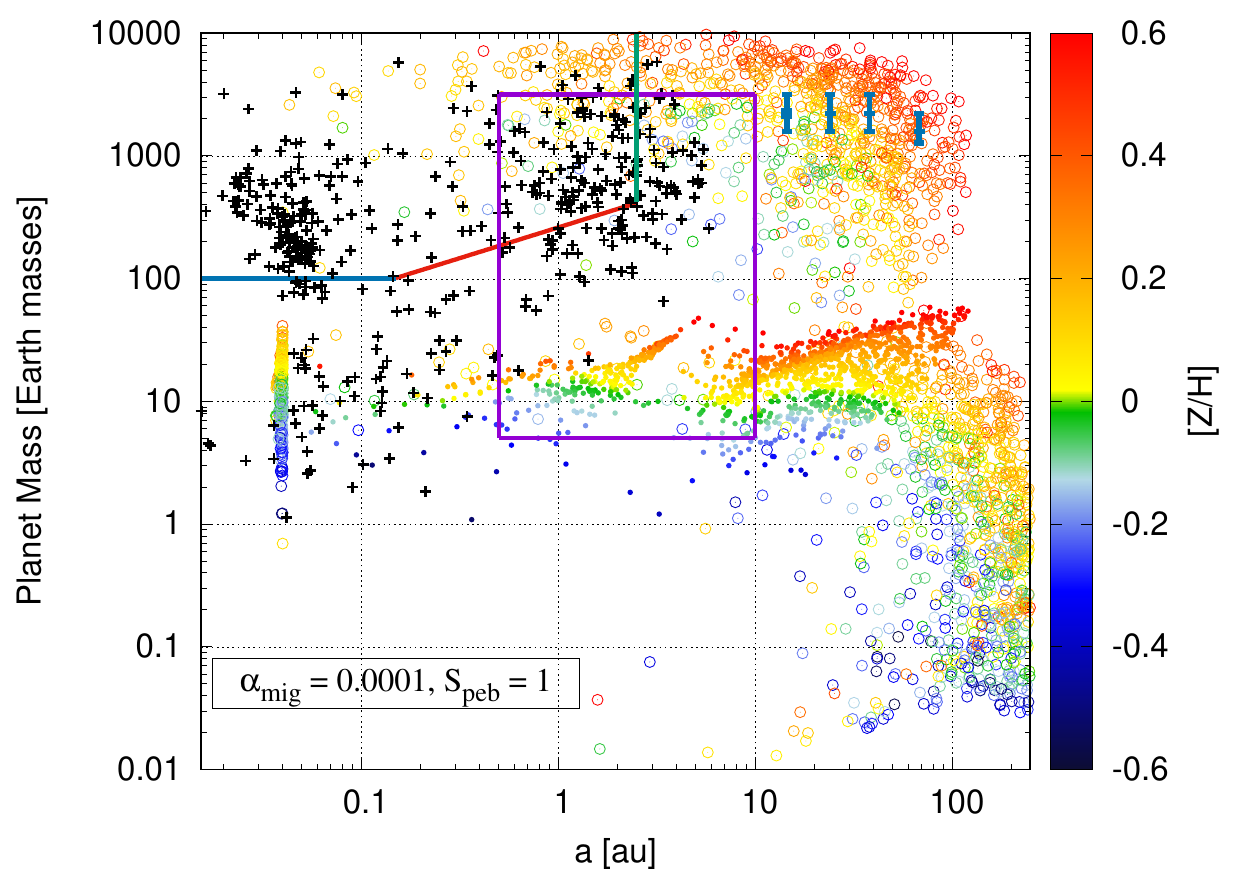}
         \hfill
         \includegraphics[width=\columnwidth, height=1.75in]{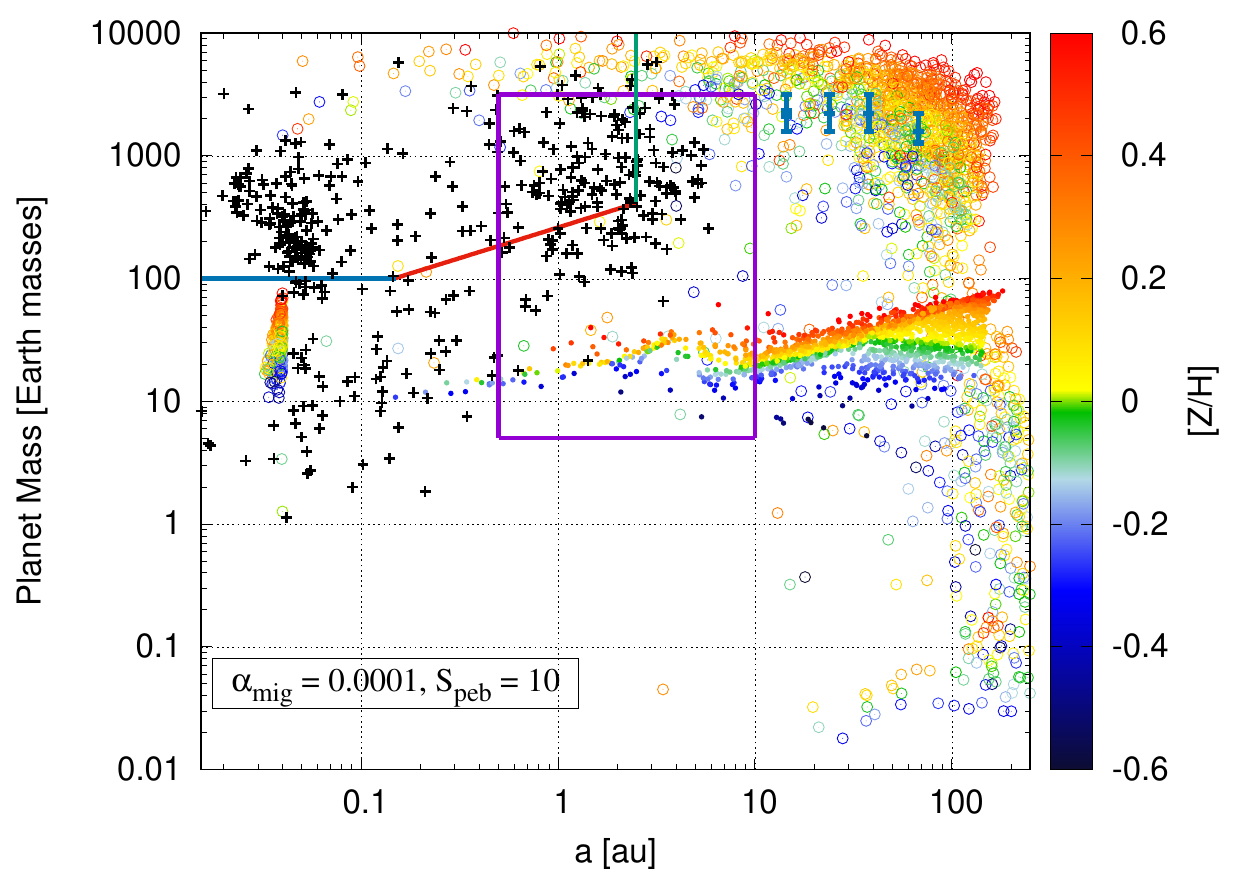}
         \hfill
         \caption{The plot has exactly the same meaning as in Fig.3 but with dots coloured as a function of metallicity. In combination with Fig.3, this shows clearly that a high metallicity is needed to reach the pebble isolation mass in the outer disc within 1-2 My (see also Fig.2).}
%          \label{fig:3}
 \end{figure}

\section{Match to radial velocity and microlensing observation}
Fig.5 shows the comparison of the synthesised gas giant sub-population to the RV measurements of \cite{Johnson2010} in a manner similar to the approach of \cite{Ndugu2018}. The plot shows that we do not produce enough gas giants compared to the RV observations for both models, even though the low alpha and high $S_{\rm peb}$ model shows a very good match to the gap statistics. One clear reason for this mismatch is the lower migration speeds which deprives the gas giants from reaching the RV field of view before the disc dissipates. We show simulations with a logarithmic starting configuration of the embryos in the appendix, which essentially recover our here discussed results. 
% We also note here, the saturation of the simulation curve at higher metallicities which is not seen in the best fit of RV curve of \cite{Johnson2010}.
% % This could possibly arise from the radial extend completeness for the RV survey therefore less synthesised data point is captured for comparison. 
% This could be due to the one embryo-per-disc limitation of our simulations which saturates the occurrence curve below the RV curve of \cite{Johnson2010}. The mentioned saturation problem could be solve by invoking N-body paradigm which our model is missing. 
To broaden the scope of our comparison, we match the synthesis data to the microlensing survey of \cite{Cassan2012}. The microlensing survey we used considered planets inside the purple rectangle in Fig. 3 and Fig.4 (see \cite{Cassan2012} for the detailed limits). From the survey in \cite{Cassan2012}, the planet frequency (for masses and distances as indicated in Fig.3 and Fig.4) scales as
\begin{equation}
 f = 10^{-0.62 \pm 0.22} \left(\frac{M_{\rm P}}{M_{\rm Sat}}\right)^{-0.73 \pm 0.17}.
\end{equation}
The fit assume a Saturnian pivot-mass (${M_{\rm Sat}}$) of 95 Earth masses. From the microlensing data, low mass planets are expected to be more common than giant planets at distances of 0.5-10 au. The fits of their survey is represented by the gray band of Fig.6. From the comparison to the microlensing survey, both models overestimate the gas giant occurrence rate and underestimated the low mass planet occurrence rates even for the model which reproduced the observed gap occurrence rates in the DSHARP campaign very well. This presents yet another direct conflict with the match to the gap occurrence rates. Testing the behavior of this conflict for logarithmic starting position distribution (see, appendix), we found the same trends and mismatch between the models in matching the observed gap positions, RV field of view and the microlensing survey. This leaves us with the following summary of our simulations: If all the gaps and rings in the DSHARP survey are caused by growing planets formed in the core accretion scenario then extremely high (even unrealistically high) amounts of pebbles are needed. This implies that either not all gaps are caused by forming planets or that we are still missing some ingredients to the core accretion scenario. For example, naturally occurring pressure bumps could help in assisting core accretion in the outer parts of the disc (e.g. \citealt{Coleman2016}).

 \begin{figure}
 
         \centering

         \includegraphics[width=\columnwidth]{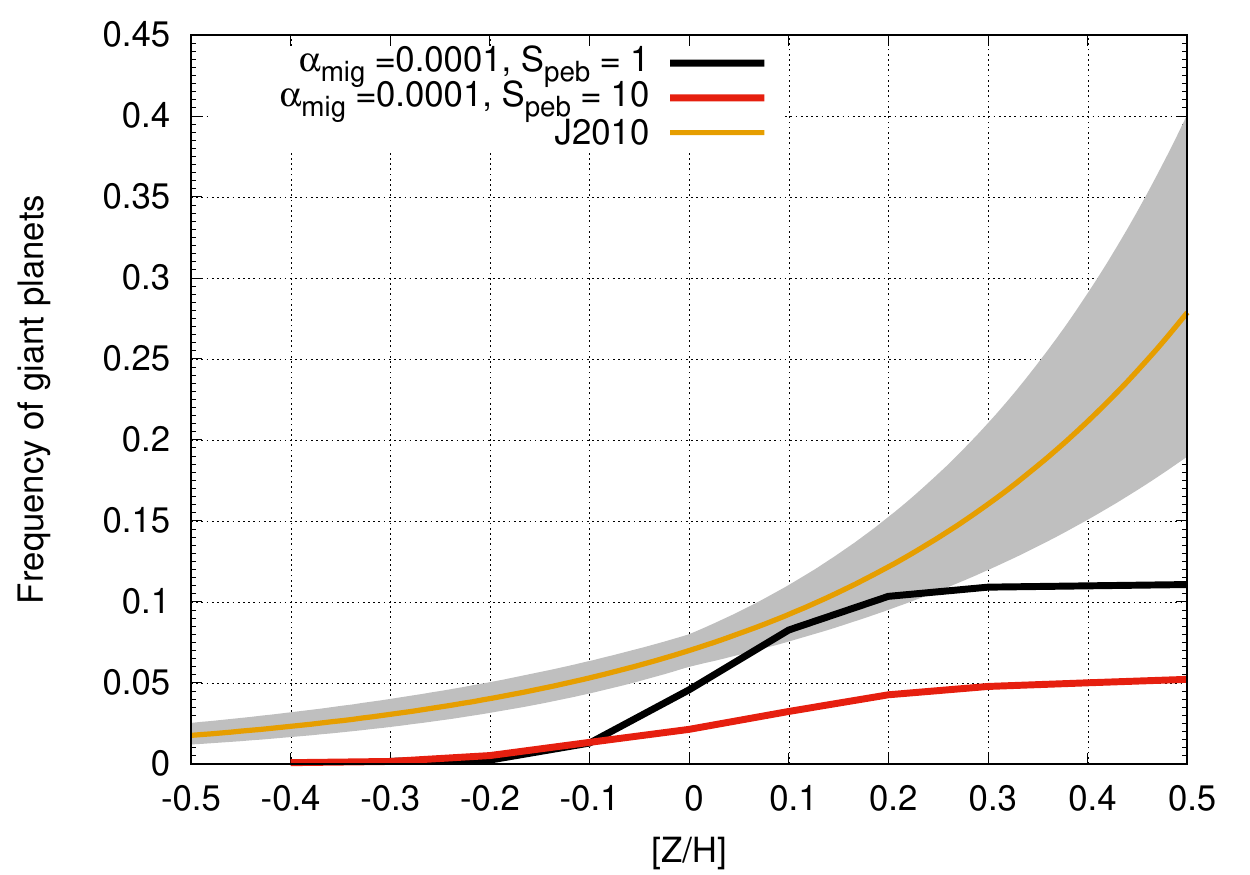}
         \hfill

 \caption{Comparison of the synthesized gas giants to the gas giant occurrence curve from the RV survey in \citet{Johnson2010}, noted as J2010 including their estimated errors. Both simulations  underestimate the giant planet occurrence rate, even though the model with $S_{\rm peb}=10$ matches the DSHARP gap occurrences (Fig.1). This is related to the fact that the planets formed mainly in the outer disc and then migrate only very slowly into the inner regions of the disc where they could be observed by RV detection.}
%   \label{fig:4}
\end{figure}

 \begin{figure}
         \centering
         \includegraphics[width=\columnwidth]{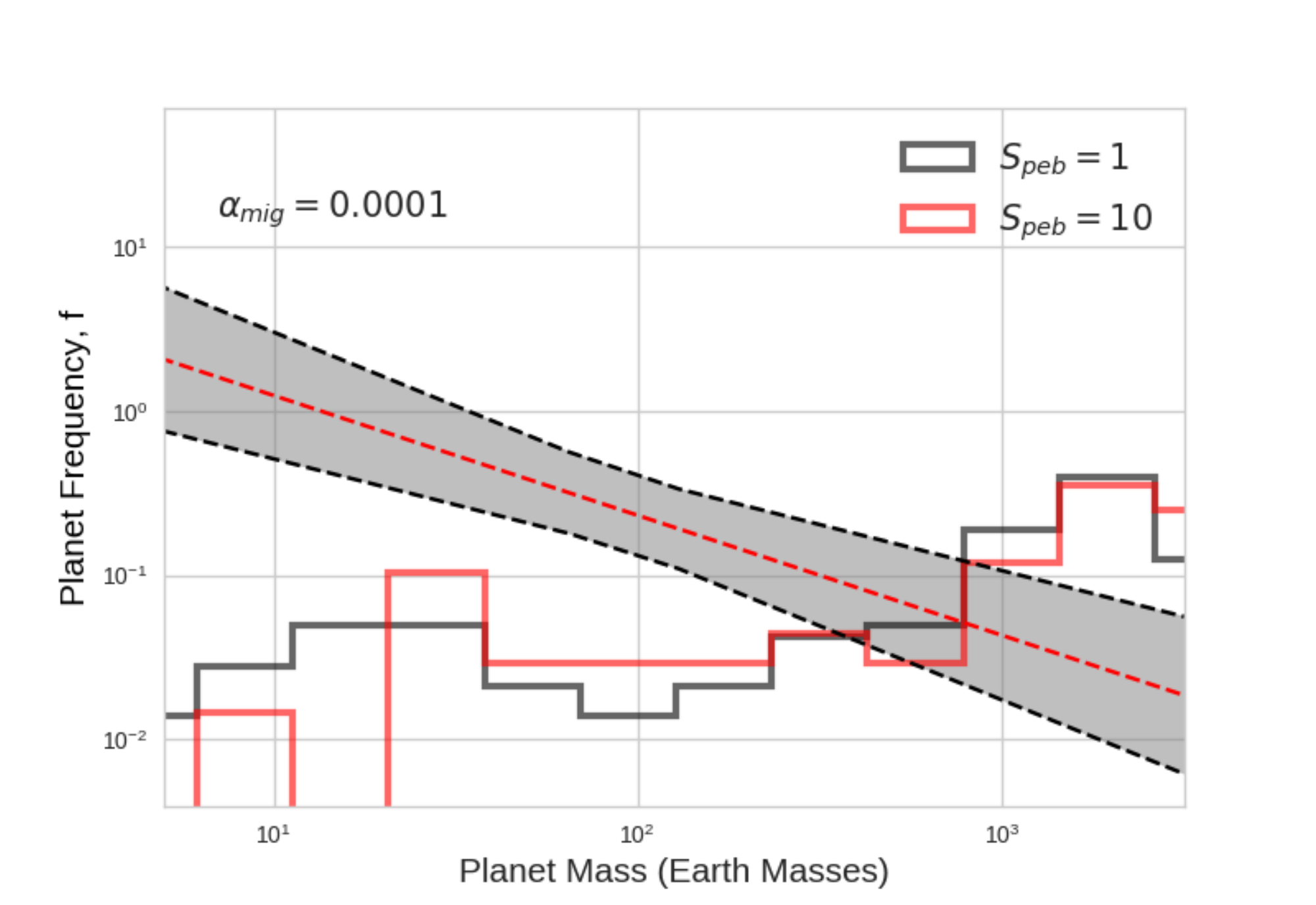}
         \hfill
 \caption{Planetary frequency determined by the microlensing survey of \citet{Cassan2012} (gray band) and our planet formation simulations. The detection of the planets correspond to the planets covered by the purple rectangle in Fig.3 and Fig.4. Our simulations seem to over predict the number of gas giants, but some of them could be removed via scattering (eventually also into the inner disc). However, our simulations clearly under predict the number of small planets in the mass regime from super-Earths all the way to Jupiter mass planets.}
  \label{fig:4}
\end{figure}

\section{Discussions}

Performing planet population synthesis from more physical photoevaporative disc models does not significantly differ from our approach. For example, in the photoevaporative disc model by \cite{Monsch2019}, the hole of photoevaporation is around 1-2 au, which is interior to the region where the planets resides at the end of the gas-disc lifetime in our simulations with low viscosity. Ideally this mechanism would help us to keep planets in the RV bin, because planet migrates less in low alpha specifications. However, we basically have no hot Jupiters that could be trapped further out by changes in the disc structure due to photoevaporation at 1-2 au. This would only play a role if the viscosity and thus the type-II migration rate is higher. At higher migration rates, it is not possible from our simulation for cores to open gaps exterior to 100 au. Therefore this is in conflict with the DSHARP survey, where gaps are also observed in the outer discs.

Our outer disc is basically not different from other detailed radiative disc models as it is determined by stellar irradiation. For example, in \cite{Baillie2015}, the outer disc structure looks very similar to our model, with H/r following a 2/7 power law with radius \citep{Chiang1997}. As the outer disc structure is similar in the disc models, because it is determined by stellar irradiation, planet formation should follow equivalent pathways.

 The only difference appears in the location of snow regions, where several snow regions were reported within 0.1-1 au in \cite{Baillie2015}. These differences arise from the different opacity model used in \cite{Baillie2015}, who used the opacities from \cite{Semenov2003}, which feature multiple condensation fronts for different chemical species, which are not included in our model.

However, the inner 1 au region is only reached in our model by gas giants, if the viscosity and thus the migration speed is high, which fails to explain the observed gap structure. In the case of low viscosity, outward migration around these ice lines would cease to exist, as the corotation torque saturates at low viscosity (e.g. \cite{Paardekooper2011}). Planets would thus also migrate inwards as in our model. This implies that even the difference in the inner structure of the disc does not have a big influence on our results.

On the other hand, if steeper and simpler power laws for the disc structure were used (e.g. an MMSN like power law disc \citep{Weidenschilling} ), planet migration would be fast and inwards in the inner regions of the disc as noted in \citet{Bitsch2015b}. However, the steeper slope of the surface density profile in the MMSN compared to our disc would greatly reduce planet formation in the outer disc, making it hard to form planets in the outer disc with this steep profile.

\section{Summary}
In this paper, we took the pebble-based planet population synthesis approach of \cite{Ndugu2018}  to simulate statistics of gaps observed in protoplanetary discs \citep{Huang2018}. The planet population model is in the one embryo-per-disc fashion where single planetary embryos evolve to pebble isolation mass where we compare our synthesized planet population with the observations of gaps in discs. The comparison of the synthesized data to the observed gap statistics (Fig.1 and Fig.2) asserts the following:

\begin{itemize}
 \item [(i)] The rings and gaps in the observed disc distributions can be caused by planets, however, if these planets form by the core accretion scenario very high (maybe even unrealistically high) amounts of pebbles are needed, especially to explain the formation of the rings exterior to 100 au.
\item [(ii)] The timescale to form planets that could correspond to the outermost gaps and rings in the observations is very close or even longer than the inferred lifetime of the observed discs. The outermost rings can only be match with very high pebble amounts and elevated metalicities, which could lead to the prediction that giant planets far way from the central star should be more frequent around high metallicity stars.
\item [(iii)] For our model to match the gap statistics observed by the DSHARP survey, a pebble production line at around 300 au would be needed. This is larger than the sizes of the observed discs. In order to match the pebble accretion scenario under the assumption that all gaps are caused by planets, the pebbles from the outer disc would have needed to drift inwards at the time of the disc observations.
\end{itemize}

In the second stage of our simulations, planet evolution is extended until the protoplanet migrates to the inner edge of the disc (0.004 au) or the end of the disc lifetime is reached. The synthetic planet population is then compared with a radial velocity observational survey \citep{Johnson2010} and a microlensing survey \citep{Cassan2012}. We clearly see from these comparisons that planets formed in the environments that match the observed gaps grow basically all to giant planets and we have no small planets left (Fig.6). This is not in agreement with observations which show that close in super-Earths \citep{Mayor2011,Mulders2018} and ice giants around the water ice line  \citep{Cassan2012,Suzuki2018} should be the most common type of planets. Our simulations thus predict, if the protoplanetary discs seen by the observations are typical discs in respect to their size and heavy element content, a large population of giant planets, which could be seen by direct imaging surveys. However, this population has not been found up to date. This leaves us with the following, not mutually excluding, conclusions:
\begin{itemize}
 \item  [(i)] Not all rings can be caused by planets, because of the contradictions to the observations.

 \item  [(ii)]  Planet formation simulations are missing some important ingredients regarding gas accretion (see also \citet{Nayakshin2019}) and planet migration.
 
 \item  [(iii)] The discs of the DSHARP survey are exceptions due to their observational bias \citep{Andrews2018} which do not represent the norm of protoplanetary discs and planet formation.

 The survey by \citet{Long2018} in the Taurus star-forming region and the survey by \citet{Cieza2019} in the Ophiuchus star-forming region, on the other hand, found a significant population of very compact discs (outer disc radii < 30 au). These discs are not pictured within the DSHARP sample, due to the selection biases in the DSHARP campaign \citep{Andrews2018}. If these discs are indeed exceptions and only 1 \% of all the disc produce these wide orbit gaps, our simulations would only apply to 1 \% of the stars and could thus represent the observed number of giant planets seen by direct imaging campaigns. 
%  \item  
\end{itemize}
We thus urge to investigate protoplanetary discs in more detail to get reliable statistics about their sizes and mass as function of their age to constrain the immediate steps of planet formation models.

\section*{Acknowledgments}
 N.N \& E.J acknowledge financial support from the  International Science Program (ISP). B.B, thanks the European Research Council (ERC Starting Grant 757448-PAMDORA) for their financial support.

 \appendix
 
 \section{Influence of the starting position of the planetary cores}
Unfortunately, no observational constraints exist about the initial distribution of planetary embryos in protoplanetary discs. To test how the initial starting position distribution of seeds affect our results, we present in this section the findings for simulations with a logarithmic distribution of the starting positions. Apart from the starting position distribution, we maintained the same parameter configurations as in our previous simulations. We use $\alpha_{\rm mig}=0.0001$ but with $S_{\rm peb}= 1$ and $S_{\rm peb}= 10$, respectively. We show the results of these simulations in Fig.A1, A2 and A3. We see that the starting configurations of the planetary embryos can have important consequences for the final configuration of the planetary systems. The logarithmic starting configuration shifts, by construction, the synthetic gap occurrence rates to the inner discs. This leads to a better match in the comparison to the RV detection of giant planets, as more planets are formed in the inner disc which can end up within the RV detection limits, compared to a linear starting configuration where most planets end up in the outer disc (Fig.3 and Fig.4). However, as stated in the main paper, scattering events if multiple planets grow in the disc could move planets from the outer population into the detection bin of RV surveys.

A comparison with the microlensing observations reveals a very similar trend as for the linear starting configurations of the planetary seeds. Our simulations over predict the amount of giant planets and under predict the number of super-Earths and ice giants. Both, the linear and logarithmic starting configuration, thus fail at reproducing the observational constraints.

The differences between the two different starting configurations are very small and do not change our comparison to observations. We thus conclude that the starting configuration does not play a major role in our assessment, but the mismatch to observations is rather explained by the conclusions made in Section 5.
 
 \begin{figure}
 \centering
 \includegraphics[width=\columnwidth]{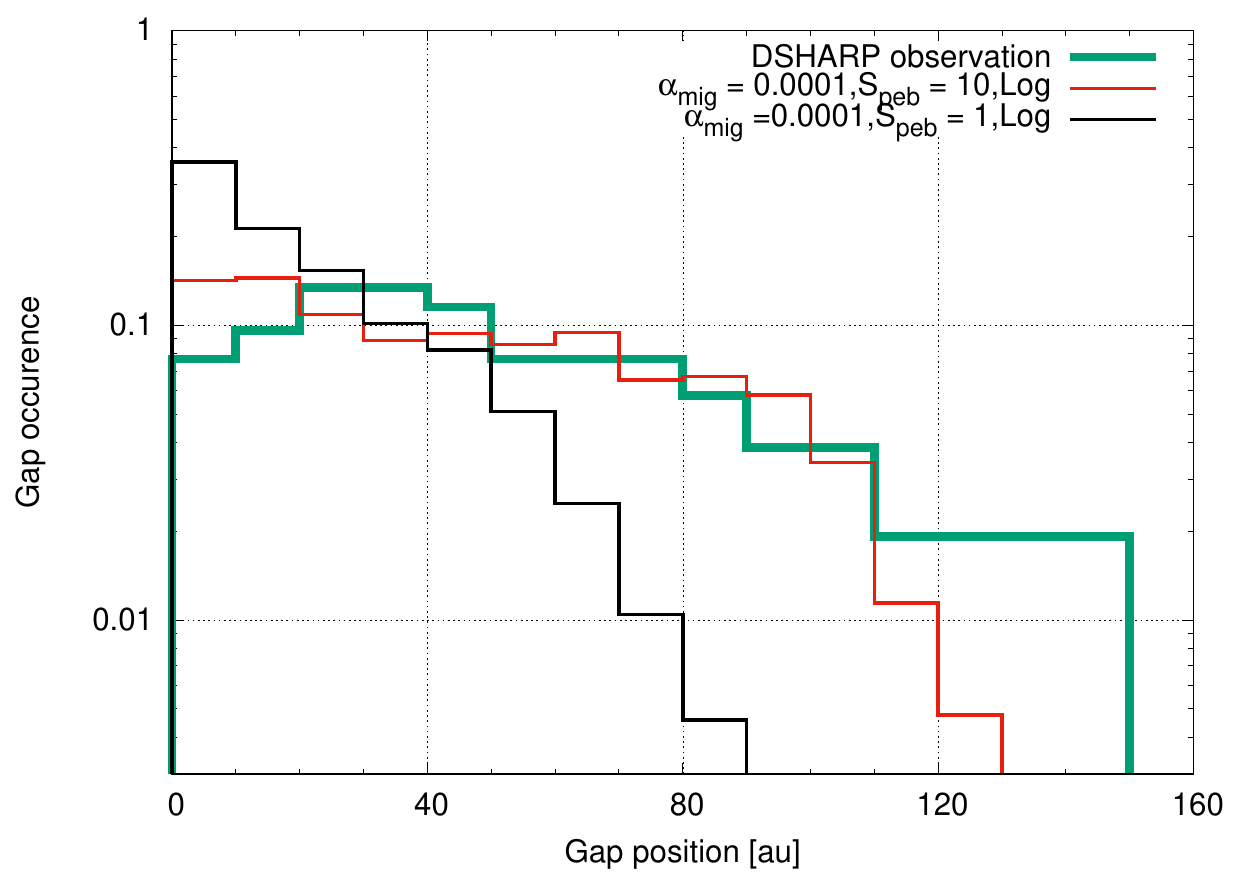}
 \hfill
 \caption{Same as Fig.1, except that a logarithmic initial semi-major axis configuration for the planetary embryos is used. The logarithmic starting configurations shifts by construction the gap occurrence rate to the inner regions of the discs, giving a bad match to the outer gap observations.}
%   \label{fig:1}
\end{figure}
 \begin{figure}
 
         \centering

         \includegraphics[width=\columnwidth]{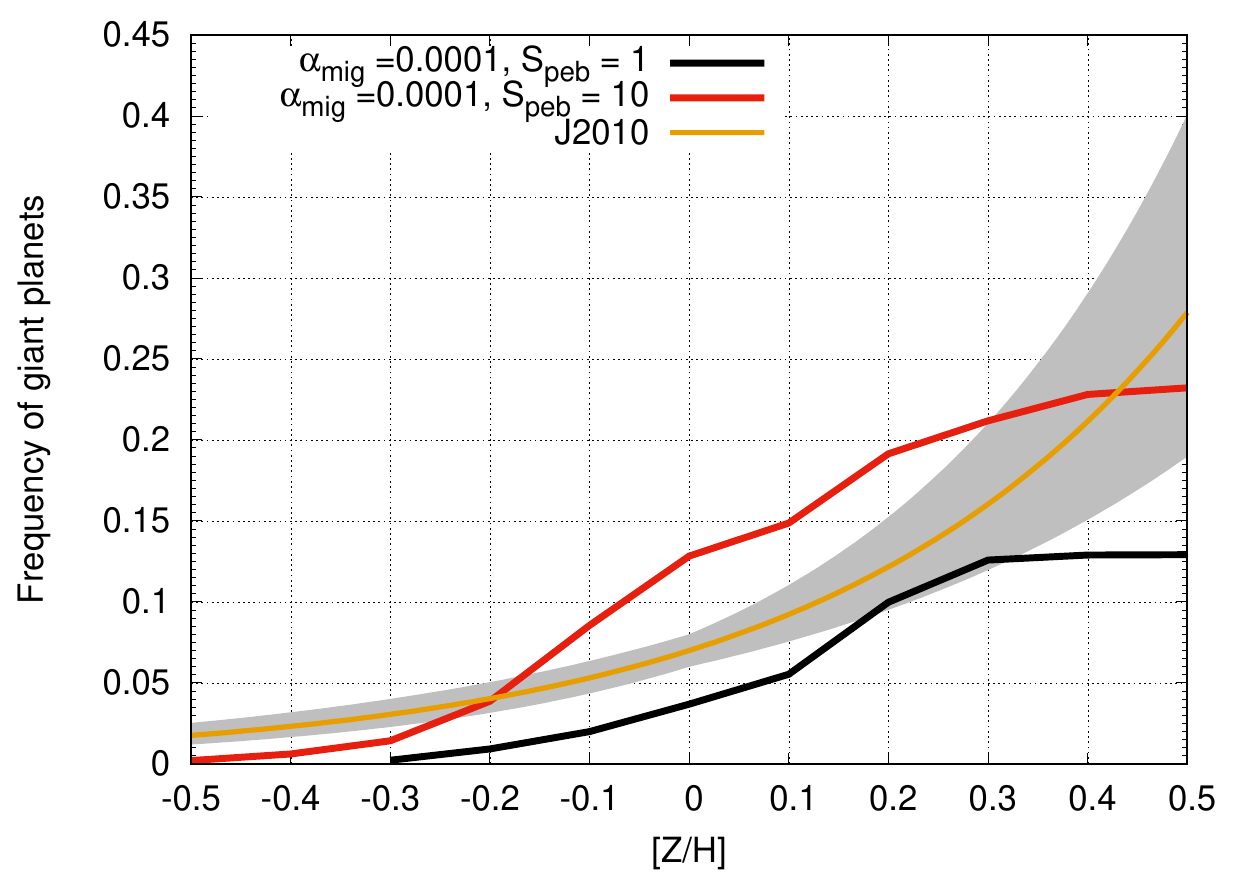}
         \hfill

 \caption{Frequency of the gas giant occurrence for the logarithmic starting configuration (Fig.5 shows a linear configuration). The model with $S_{\rm peb}=10$ shows a good comparison to the RV data from \citet{Johnson2010}.}
%   \label{fig:4}
\end{figure}

 \begin{figure}
         \centering
         \includegraphics[width=\columnwidth]{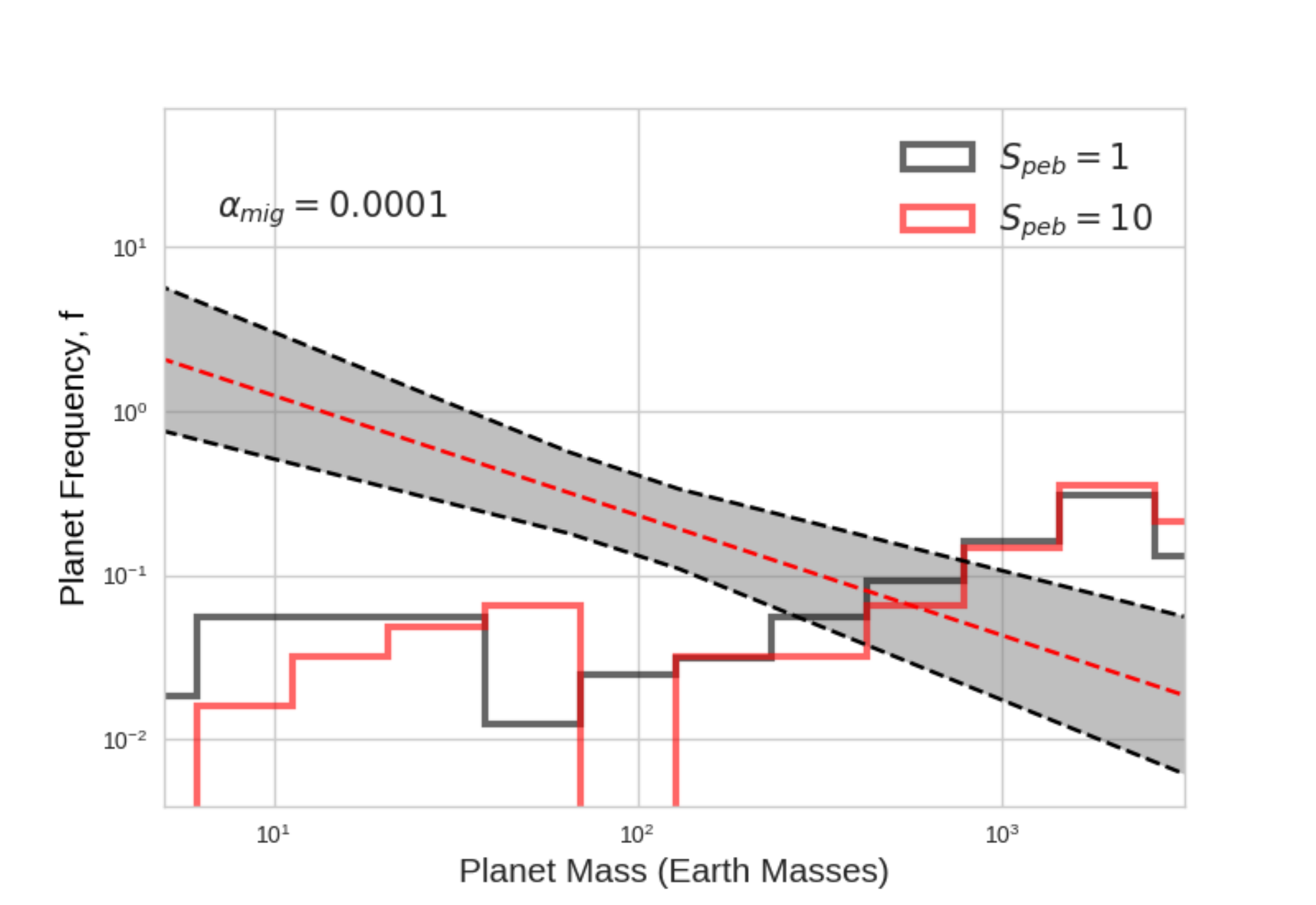}
         \hfill
 \caption{Same as Fig.6, except that a logarithmic starting configuration for the initial semi-major axis of the embryos is used. The results indicate that, as for the linear starting configuration, gas giants are overproduced and super-Earths and ice giants are under produced in our model, showing a clear mismatch to observations.}
%   \label{fig:4}
\end{figure}

\bibliographystyle{mnras}
\bibliography{biblio.bib}

% \label{lastpage}

% Don't change these lines
\bsp	% typesetting comment
\label{lastpage}
\end{document}